# Analysis and correction of errors in nanoscale particle tracking using the Single-pixel interior filling function (SPIFF) algorithm

Yuval Yifat[1], Nishant Sule[1], Yihan Lin[3], Norbert F. Scherer[1,2,*]

[1] *James Franck Institute,* [2] *Department of Chemistry, The University of Chicago, Chicago Il. 60637, USA.*

[3] *Center for Quantitative Biology, Peking-Tsinghua Center for Life Sciences, Academy for Advanced Interdisciplinary Studies, Peking University, Beijing 100871, China.*

\* Corresponding Author: nfschere@uchicago.edu

**Abstract**: Particle tracking, which is an essential tool in many fields of scientific research, uses algorithms that retrieve the centroid of tracked particles with sub-pixel accuracy. However, images in which the particles occupy a small number of pixels on the detector, are in close proximity to other particles or suffer from background noise, show a systematic error in which the particle sub-pixel positions are biased towards the center of the pixel. This "pixel locking" effect greatly reduces particle tracking accuracy. In this report, we demonstrate the severity of these errors by tracking experimental (and simulated) imaging data of optically trapped silver nanoparticles and single fluorescent proteins. We show that errors in interparticle separation, angle and mean square displacement are significantly reduced by applying the corrective Single-pixel interior filling function (SPIFF) algorithm. Our work demonstrates the potential ubiquity of such errors and the general applicability of SPIFF correction to many experimental fields.

**Introduction:**

Imaging has become an increasingly important part of scientific research. Breakthroughs in biology[1], material science[2,3] and astrophysics[4] have been made possible by advances in imaging systems[5] and techniques that allow rapid measurement of physical phenomena below the resolution limit[6–9]. In an optical imaging system, light from an object of interest is focused onto a detector and converted to electrons that are digitized and stored as a two-dimensional array (of pixels) representing the position-dependent intensity map in space (and a third dimension as a video, in time). Once the image is obtained, mathematical algorithms are used to determine the particle positions in each frame, and to track them. One needs to establish the unique identities of the individual particles across frames, to create trajectories.

Many particle tracking algorithms use the distribution of pixel intensities along with knowledge about the point-spread function (PSF) of the system to localize the particle with sub-pixel accuracy. Widely used ones include: the "Crocker-Grier" algorithm[10], which determines the "center of mass" of pixel intensities to estimate the location of the particle; the Raghuveer algorithm[11], which calculates the maximum radial gradient around the particle to estimate its center; and non-linear fitting of a Gaussian function to the pixel intensity distribution[12,13]. However, despite continued development of algorithms and benchmark comparisons between them[14–16], problems still arise with their accuracy and efficacy, especially when images suffer from low signal-to-noise, are cluttered with multiple objects in close proximity, or are of particularly small objects.

One such problem is that all algorithms, to a greater or lesser degree, bias the sub-pixel location of a tracked object towards the center of the calculated pixel. This effect is known as "pixel locking" or "pixel biasing"[17–19]. Recently, Burov et al[20] reported the Single Pixel Interior Fill Function (SPIFF) algorithm that corrects this bias even in cases where the actual tracking is done with a proprietary method and is unknown. The correction of the pixel locking error is done by analyzing a set of images, collecting the decimal part of the tracked locations in a "meta-pixel", and determining if the resultant distribution is uniform within the meta-pixel. The SPIFF algorithm uses this distribution to estimate the true particle position by expanding the probability distribution such that it uniformly spans the entire meta-pixel. Thus, given a set of estimated sub-pixel values of particle centers $\{\hat{X}_E\}$, the true particle position, $\hat{X}_T$, can be calculated by numerically solving the integral[20]

$$\hat{X}_T = \pm \int_0^{\hat{X}_E} P(\hat{X}_E') d\hat{X}_E', \qquad (1)$$

and finding the fraction of estimated values in the range $(0, \hat{X}_E)$, where $P(\hat{X}_E')$ is the SPIFF density function, i.e. the probability density of the set $\{\hat{X}_E\}$ (see ref [20] and Supporting Information).

In this report we establish the importance of the SPIFF algorithm by demonstrating the prevalence of pixel locking errors in common experimental systems (beyond the colloidal particles considered in[20]), their detrimental effect on experimental outcomes, and how they can be identified and corrected. We study tracking errors of nanoparticles and single molecules *and* demonstrate quantitatively the consequences of the error correction that the SPIFF algorithm provides vis-a-vis its effect on statistical and dynamical properties of these systems. We demonstrate the severity of errors that can arise from biased tracking for common experimental conditions when the size of the particles (and/or the tracking window) is small (i.e. occupies two to four pixels on the detector), and the sampling rate is high (i.e. particle motion is sub-pixel per frame). We also illustrate how the SPIFF algorithm ameliorates these errors and greatly improves the fidelity of experimentally extracted quantities such as particle trajectories or mean square displacement (MSD). We also show limitations of a "global" SPIFF correction and offer additional strategies for 2nd tier error correction in the Supporting Information. Overall, we demonstrate pixel locking error resulting from common biases inherent to a wide range of experimental results can be easily identified and now corrected.

**Results:**

**Nanoparticle imaging:**

Since nanoparticles are often smaller than the PSF of optical microscopes and thus can be separated by sub-resolution limit distances, errors in their localization can be severe. Therefore, we considered the general problem of tracking several closely spaced nanoparticles. We analyze experimental data obtained by imaging 150 nm diameter Ag nano-particles that are trapped and manipulated using holographic optical tweezers. Our experimental setup is presented in the Supporting Information. Optical tweezers have been used in a wide range of research fields as they allow the manipulation of micro and nanometer sized objects[21,22] in order to understand interparticle dynamics and behavior[22] or light-mater interactions[23]. The goal of our experiment is to understand the forces and torques exerted on collections of nanoparticles for different polarization states of the trapping beam. Therefore, extracting accurate interparticle separations from tracking particle locations is crucial.

As shown in Figure 1, which is a representative frame taken from such an experiment, two or more particles can be confined in the electrodynamic near-field, i.e. they are separated by roughly 200 nm center-to-center making particle tracking difficult due to the overlap of the individual particle images on the detector. The size of the nanoparticles and their proximity necessitates that the window function used for tracking only be a few pixels wide to avoid misidentification of 2 particles as a single one. Such small widows manifest under-sampling errors and cause strong pixel locking[20]. To reduce or avoid pixel locking, one can increase the size of the tracking window, $W$ used for fitting to the (expected) Gaussian intensity distribution. This could indeed work when the particles are large (compared to the PSF) or are well separated, but when small particles are in close proximity their images overlap. Thus, the particle center locations are necessarily in error, and an increasing fraction of the particles are misidentified. Misidentification (and undercounting) is not an error that can be corrected by the SPIFF algorithm. More specialized methods that model multiple particles with prior information might be used instead[24]. Herein, we deal with pixel locking error and SPIFF correction.

The tradeoff between tracking window size and misidentification is demonstrated in Figure 1, where 1000 frames taken from an experimental video of the Ag nanoparticles were tracked using the Mosaic tracking software for ImageJ[25], which uses nonlinear least squares Gaussian fitting for centroid determination. In Figure 1 we used "circular" windows with radii $R=1,2,3$ and 7 pixels (window size $W$ defined as $W=2R+1$). Column 1(a-d) in the figure shows representative images of the localization for each window size while column 1(e-h) shows the corresponding distributions of sub-pixel localizations in the meta-pixel. In addition, we counted the number of particles found for each window radius and calculated the percentage of frames in which 3 particles were identified to be 93%, 83%, 58% and 33%, respectively.

The compression of the density toward the center of the meta-pixel in Figure 1(e-f) is (a signature of) pixel locking error created by a small window; i.e., a Nyquist sampling error or bias[20].

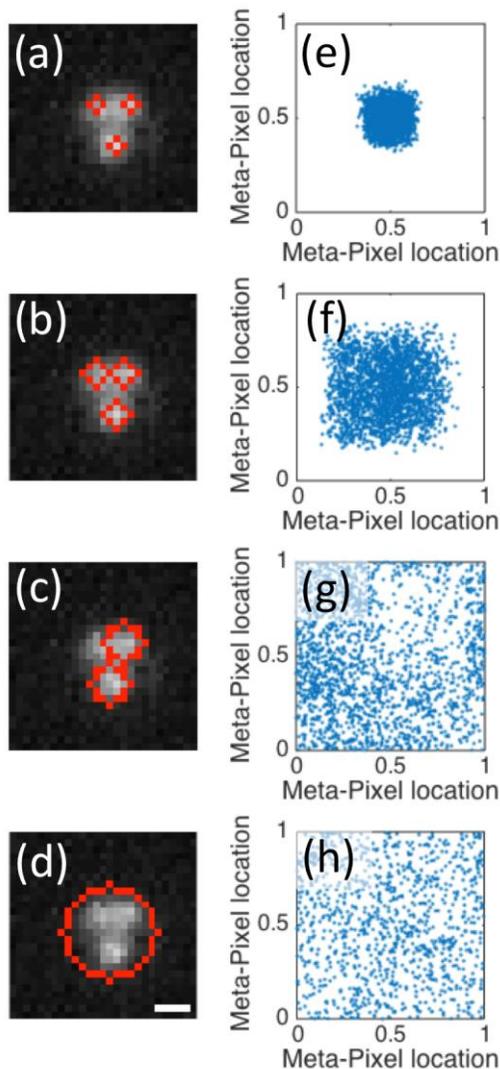

**Figure 1**. The effect of window size, *W*, or windowing on the particle identification (left column) and the associated meta-pixel distributions (right column). Images (a) – (d) show a representative frame taken from a video of three 150 nm diameter Ag nanoparticles trapped in a linearly polarized Gaussian trap. The scale bar represents 360 nm. The particles in the frame were identified by running the Mosaic algorithm described in the text with window radius values of 1 pixel (frame a), 2 pixels (b), 3 pixels (c) and 7 pixels (d). Panels e-h show the corresponding sub-pixel localization distributions obtained from analyzing $10^3$ frames of a video of 3 nanoparticles in an optical trap. The experimental magnification of 90x gives an effective pixel size of 72 x 72 nm.

When a small window size is used, almost all the particles are identified and tracked in each frame, but the pixel locking error is pronounced and any metric obtained from the particle tracking (distance, angle, MSD etc.) will be biased. The errors are less pronounced for larger window sizes, but at the cost of misidentifying (and losing) particles. Taken ad absurdum, we reach the results shown in Figure 1(d,h); when the window is large enough to encompass all three particles, the tracking algorithm treats them as a single entity and the pixel locking error vanishes. However, the tracked "particle" clearly has little to do with the actual experimental results.

**Tracking Synthetic experimental data:** Since it is impossible to know the "true" positions of the tracked particles in an optical trapping experiment of mobile particles, we simulated the results of such an experiment and applied the tracking algorithm to create a benchmark for the SPIFF correction efficacy. This was achieved by simulating frames of two silver particles at varying separations based on experimental videos (such as that described in[22]). We used typical background intensity and variation and particle intensity and size to create frames of synthetic data (images) that simulate the intensity distribution on the detector for given particle locations. The procedure for image synthesis is given in the Supporting Information. Representative examples of experimental and synthetic images of two 150 nm Ag particles are shown in Figure 2(a,b).

Using this method, we synthesized images of nanoparticles at arbitrary separations and investigated the results of particle tracking. The procedure to create synthetic data was applied to a list of $10^4$ positions in which the first particle was fixed at the center of pixel (0,0) while the second one was randomly positioned around (5,0) according to a normal distribution with $\sigma_x=2$ and $\sigma_y=1$ pixels (see Supporting Information). For each frame, we tracked the positions of the particle using the Mosaic algorithm with window radii of $R=1$ (essentially a Swiss cross shaped particle identification window), and 2. The tradeoff described earlier was inherent in this analysis as the Mosaic algorithm correctly identified two particles in 84% of all frames when we used a window with $R=1$, compared to 79% of all frames with $R=2$.

The pixel locking bias is evident in Figure 2(c), which shows the original distribution of particle 2 as well as the tracked positions for all frames where two particles were identified (neglecting cases where the particles were too close to be visually separated). As the tracking algorithm locks the particle location towards the center of a pixel, the distribution of particle positions changes from a Gaussian (blue dots) into sub-pixel regions on a two-dimensional lattice (red and pink sub-pixel size squares). It is evident from this figure that a smaller tracking window compresses the tracked distribution towards the center of the meta-pixel, as shown in Fig. 1. In addition, note the absence of pink dots in the areas where the interparticle separation was small. The fact that only red dots ($R=1$) are observed for those small separations means that the algorithm only succeeds in identifying two particles for the smaller value of $R$.

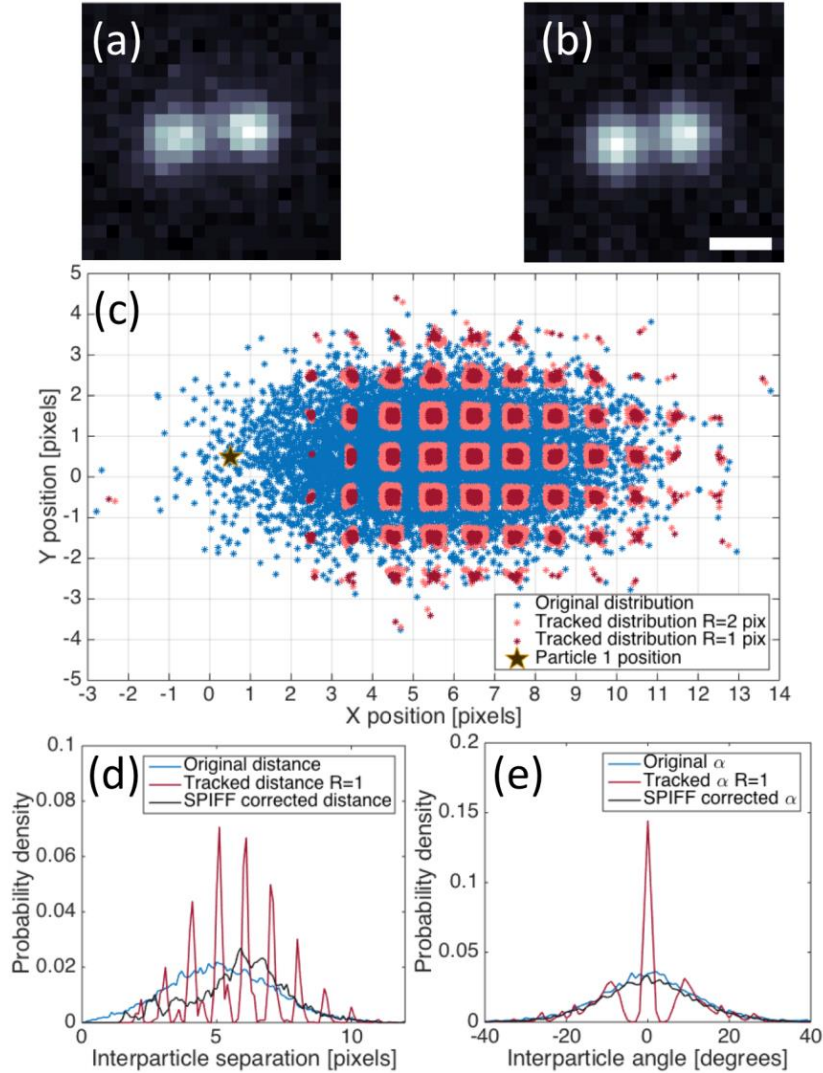

**Figure 2** – Representative examples of experimental (a) and synthetic (b) images of two particles separated by 440 nm. Scale bar represents 360 nm. Details of the experimental setup and procedure for creating synthetic images are described in main text and Supporting Information. (c) Distribution maps for particle 1 (marked as black star) and true positions for particle 2 (blue dots). The associated synthetic images were tracked using the Mosaic nonlinear least-square Gaussian fitting algorithm as described in the main text. The red and pink dots show the distributions of tracked particle positions with windowing radii of R=1 (red) and R=2 (pink), respectively. Note that particle 1 was positioned at (0.5,0.5) which is the center of pixel (0,0), the definition of a pixel location in the Mosaic algorithm. (d) Probability density functions of Cartesian interparticle separation and (e) angle for the original data (blue) as well as the tracking algorithm biased (red) and SPIFF corrected values (black). Only frames in which both particles are identified are considered in the distribution map and the probability density histograms. As a result, there are fewer measurements in the tracked and SPIFF corrected distributions in panels (c,d) and these distributions are normalized relative to the true particle count. The normalization is done to avoid artificially boosting the probability distribution values at larger separations where both particles are consistently identified.

We corrected the pixel locking bias using the SPIFF algorithm given by Eqn. (1). Figure 2(d,e) shows the interparticle angles and separations – two metrics commonly used in particle dynamics analysis, generated from the tracked and corrected localizations rate. As evident from Fig. 2(d), the calculated interparticle separation is strongly affected by the bias caused by small window sizes, due to a Nyquist sampling error. Since the tracked positions are locked to the centers of the pixels, the calculated Cartesian separation becomes discontinuous, evident as red spikes. The SPIFF corrected localizations, represented by the black distributions in the panels, alleviates the pixel locking error. The correction works better for larger separations than for smaller ones.  One reason for this is that the intensity distributions overlap more strongly for small separations, thereby decreasing the accuracy of centroid determination[20]. Additionally, the dearth of points at smaller separations means that the correction obtained from solving Eqn. (1) underrepresents these points and is skewed. We discuss these issues in the Supporting Information.

The calculated interparticle angle, presented as the red distribution in Fig. 2(e), shows another striking error. The angles around $\pm 5^0$ are "forbidden" owing to the discontinuous nature of the pixel locked particle positions. These angular errors are corrected by applying the SPIFF algorithm to the tracked results; the correction is evident by comparing the blue and black distributions that represent the original and SPIFF-corrected interparticle angles, respectively. The improvement in the distribution of angles is better than for the distribution of separations because the SPIFF correction "fills in" all the discontinuities, thereby recovering values of angles that were missing in the original tracking distribution. By contrast, while the SPIFF correction improves the distribution of interparticle separation, it cannot recover instances when the particles are too close together and are not individually identified. Thus, most of the frames in which the interparticle separation is smaller than roughly 5 pixels (i.e. ~360 nm), are not identified, and are therefore unrepresented in the distribution in Fig. 2(d).

**Analysis of simulated particle trajectories:** Next, we analyze the results from a physically realistic simulation based on a combination of finite difference time domain (FDTD) electrodynamic simulations with Langevin dynamics (termed ED-LD). These particle trajectory data from ED-LD simulations allow analyzing the bias and SPIFF correction when the probability distribution of particle positions matches experimental results[26]. The ED-LD simulation data also allows exploring the MSD of a particle. We simulated the trajectories of two 150 nm diameter Ag particles trapped in a focused Gaussian beam linearly polarized along the x axis using methodology and parameters described previously[26]. We synthesized images of the particles at these simulated positions that were then tracked. The resultant trajectories consist of 3,200 frames with a time step of *0.5μs*.

Figure 3 shows the trajectories of two electrodynamically interacting particles and the analysis of the effects of tracking and SPIFF correction (video given in Supporting Information). Fig. 3(b) shows the original trajectory of the particle on the left (in red) as well as its tracked positions, which are the blue dots that are biased towards the pixel centers (i.e. the particle localizations are restricted to the centers of a few pixels). It is clear from the biased distribution that much information about the actual particle dynamics is lost. By contrast, the SPIFF-corrected trajectory (black) in Fig. 3(b) is a high-fidelity reconstruction of the original particle path. The fidelity of the SPIFF-corrected data is further demonstrated in Fig. 3(c-d), which shows the RMS error, defined as the square root of the Cartesian difference between the particle's true position and its tracked position, before (blue and red) or after (green and purple) SPIFF correction. Fig. 3(c) shows the time evolution of this error throughout the first 150 frames of the simulation and Fig. 3(d) shows their associated probability densities over the entire simulation. The mean error for the tracked data (using the Mosaic program) are 0.3 and 0.19 pixels with standard deviations of 0.13 and 0.09 for windows of radius *R=1* or *2*, respectively. After applying the SPIFF correction algorithm this was improved to 0.11 with a standard deviation of 0.05 pixels for both tracking window sizes (i.e. *R=1,2*).

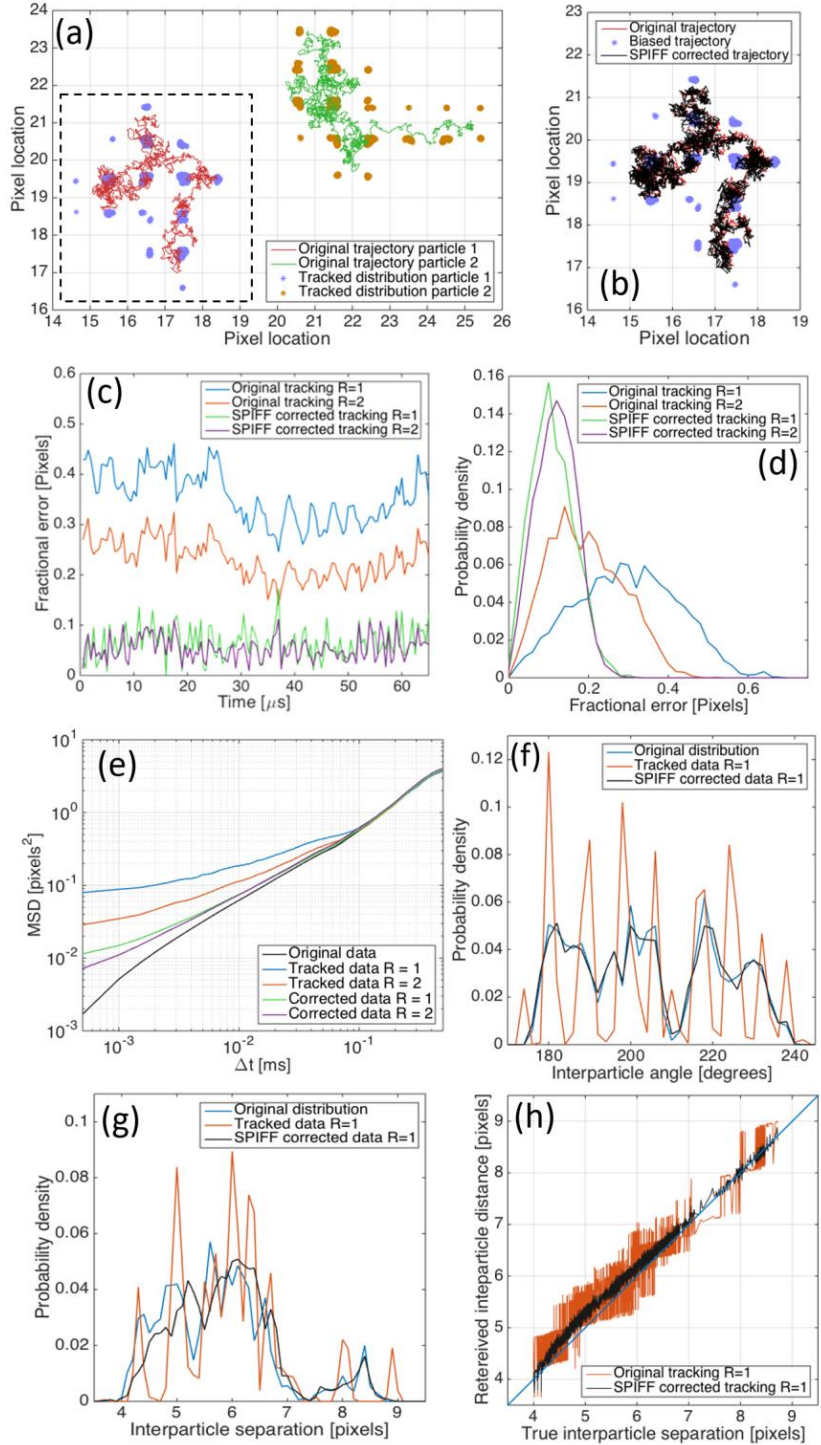

**Figure 3** – Results from tracking analysis ED-LD simulations of Ag nanoparticle pairs. Images "synthesized at the positions obtained from the simulation described in main text and Supporting Information. (a) Original particle trajectories for particles 1 (red) and 2 (green) from simulations. The associated localizations from tracking the synthesized images using the Mosaic program with a window of radius *R=1* (blue, orange dots) exhibit severe pixel locking. Pixel size is 72 nm. (b) The trajectory recovered for particle 1 (dashed rectangular box in panel (a)). Red

connected points are the original particle trajectory, blue dots are the tracked positions and black connected points are the trajectory obtained after SPIFF correction. (c) RMS trajectory error for the first 150 frames of the simulation. The results shown are the RMS error for frames tracked with the Mosaic algorithm and a window of radii *R=1,2* before and after SPIFF correction. (d) RMS error distribution for the entire tracked trajectory shown in panel (b). (e) Logarithmic plot of mean square displacement (MSD) of particle 1 as a function of time step Δt (i.e. inverse frame rate) for original, tracked and corrected trajectories. Calculated interparticle angle (f) and separation (g) probability densities (using a window of *R=1*) and the SPIFF corrected data (black). Note that the accuracy of the separation calculation is reduced for smaller separations as explained in the main text. (h) Calculated interparticle separation for the tracked (red) and SPIFF corrected (black) trajectories as a function of the true interparticle separation. Blue line is x=y.

The SPIFF correction algorithm allows more accurate measurement of physically significant properties such as the mean square displacement (MSD) of a particle, defined as:

$$MSD(\tau) = \overline{(\vec{x}(t+\tau) - \vec{x}(\tau))^2}, \qquad (2)$$

where $\vec{x}(\tau)$ is the position of a particle at a time $\tau$ and $\vec{x}(t+\tau)$ its position after time *t*. This measure is used to ascertain the nature of particle motion in conjunction with particle tracking algorithms for a wide range of fields including biophysics[27] and fluid dynamics[28]. The strength of the SPIFF correction algorithm is evident in Figure 3(e), which shows that the error of the SPIFF corrected MSD is reduced by almost an order of magnitude relative to the tracked data, increasing the fidelity of the measured results to the original trajectory and dynamics.

Analysis of interparticle separation and angle are shown in Figure 3(f,g). Similar to Figure 2, we observe significant errors, resulting from discontinuities in the probability distributions that are ameliorated by the SPIFF correction. The correction is better at larger interparticle separations (>5 pixels) than at smaller ones because, as explained previously, the tracking algorithm has difficulty identifying two particles at small separations, and the meta-pixel distribution is increasingly skewed from the pixel center as the separation decreases. This is evident in Fig. 3(h), which shows the relationship between the true interparticle separation (black connected dots) and the separation calculated from both the tracked (red) and the SPIFF corrected data (black, using a window with *R=1*). As shown in the panel, the error in the calculated separation is not constant but varies with true interparticle separation. As is discussed in the Supporting Information, this separation-dependent error can be applied in a second correction step that further increases the accuracy of interparticle separations.

**SPIFF correction of experimental data:** With these insights from simulated data, we applied the SPIFF correction algorithm to experimental data in which three Ag particles were trapped in a linearly polarized Gaussian trap. Due to the tightness of the trap and electromagnetic interactions between the particles, two particles were trapped in close (near-field) proximity near the center of the focused optical beam with a center–to-center separation of approximately 250 nm (~3-4 pixels), while the third particle was trapped close to the optical binding distance[22,29,30] with a center-to-center separation of around 470 nm (~6-7 pixels). We captured images at 1000FPS and analyzed 780 frames. Only frames where all three particles were identified were taken into consideration (video S2). Figure 4 shows a representative frame from the video as well as the distributions of particle localizations obtained using the Mosaic program with *R=1* window. The distribution of tracked particle positions exhibit significant pixel-locking.

Figure 4(b,c) shows the same errors that are apparent in our previous analyses of particles at small separations. The distribution of interparticle angles, $\theta_{12}$ (magenta curve), exhibits a similar angular profile to those observed in the simulated data in Figures 2(e) and 3(g). The omission of some angles for particles in close proximity is entirely the result of pixel-locking. The SPIFF correction recovers a continuous angular distribution as is appropriate for Brownian motion. The simulated errors are also observed for the interparticle distance $d_{12}$. Note that the distribution of errors in angle and separation are less prominent, and the SPIFF correction is milder, for particles with larger separations, such as $d_{13}$ (cyan curves). However, pixel-locking exists even for particle 3 as shown in Fig. 4(a).

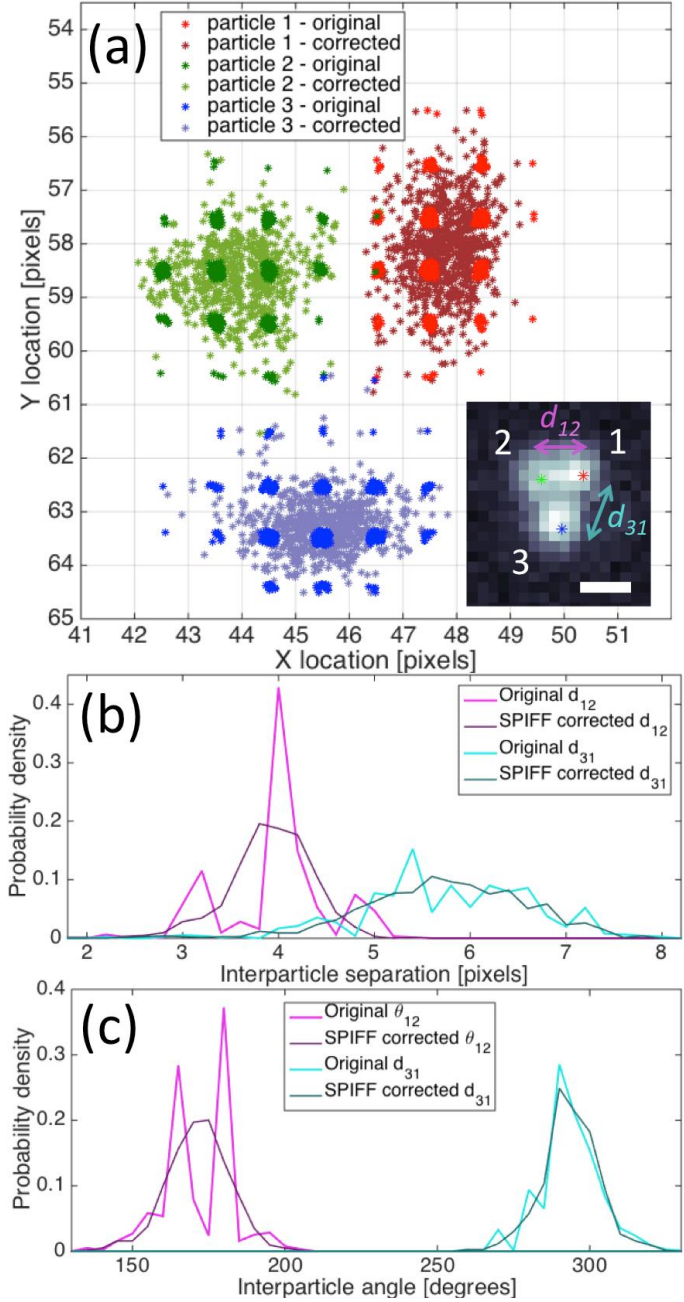

**Figure 4** Analysis of an experimental video of three particles trapped in a linearly polarized Gaussian trap. Pixel dimensions are 72 nm. (a) Particle localizations from tracking before and after SPIFF correction for the three particles. The red, green and blue marked particles are defined as particles 1,2 and 3, respectively. Inset - Representative frame from the video (full video given in SI, scale bar represents 360 nm). Interparticle separation (b), and angle (c) for the three particles. The lighter and darker connected points represent the original and tracked SPIFF-corrected values, respectively. Note the differences in the shapes of the distribution before SPIFF correction between the near-field bound particles (magenta connected points - $d_{12}$) and the optically bound ones (cyan - $d_{31}$).

**Pixel locking errors and SPIFF correction of single molecule experiments:** Determination of the positions and dynamics of single molecules has become an essential experimental method in biophysics and cell biology[31,32]. The small size, low to moderate SNR of the data (i.e. low counts and/or significant background noise) and potentially crowded nature of single molecule images can all result in tracking errors manifest as pixel-locking. We reviewed the data previously published by Lin et al[33] in which proteins were labeled with fluorophores and imaged as they traveled along flow-extended double-stranded DNA under several flow conditions. The MSD values reported in the paper were obtained from tracking video data using the Gaussian fitting method in DiaTrack 3.0[34] software suite.

We re-analyzed the reported particle trajectories and found pixel-locking in the trajectory data (Fig. 5(a)). Surprisingly, the SPIFF correction did not improve the results. We attribute this null result to the low frame rate (20FPS) and the associated large displacements of the single molecules in 50 msec. That is, the MSD values obtained (Fig. 5(b)) are at least two orders of magnitude larger than the pixel locking error (Fig. 5(c)).

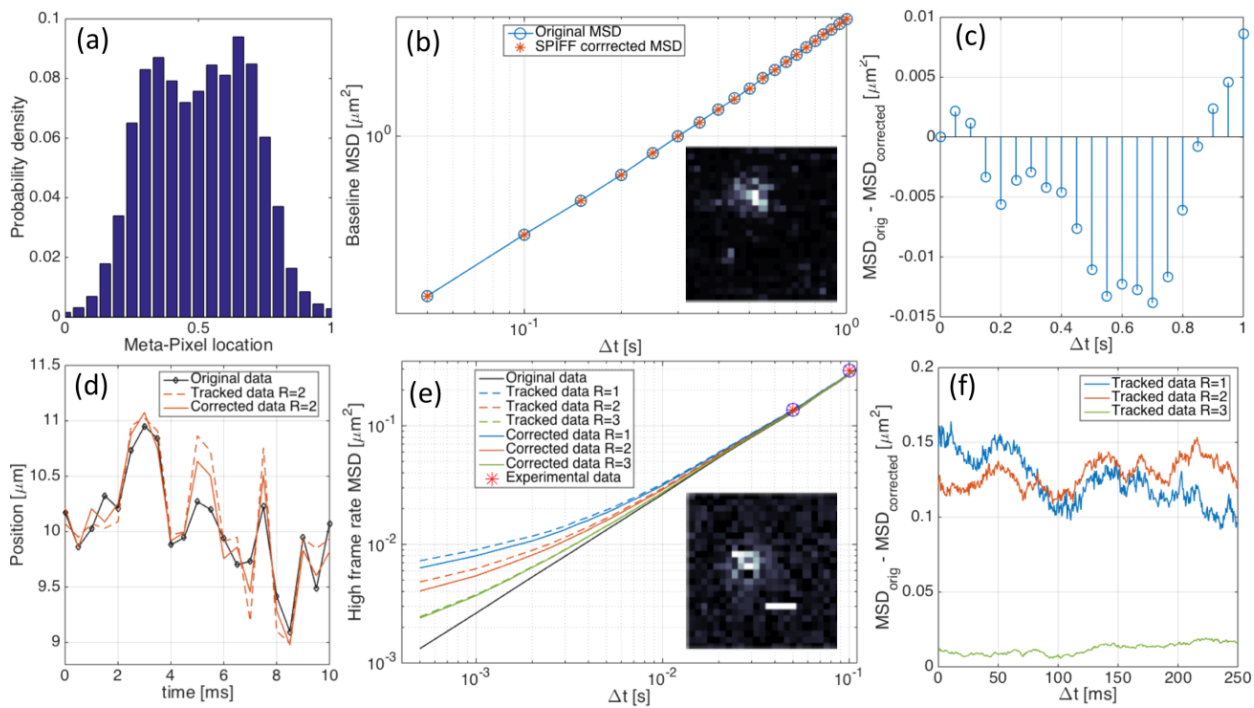

**Figure 5** – Analysis of tracking data of single molecules moving along a flow-extended ds-DNA as explained in the text. (a) Sub-pixel localization distributions obtained from the x-coordinate of the tracked single molecule positions. The non-uniformity of the distribution is pixel-locking error. (b) Original and SPIFF corrected MSD values based on trajectories reported by Lin et al[33]. Inset shows a representative experimental image. (c) Difference between original and SPIFF corrected MSD as a function of time difference. Note that there is no clear trend and the difference is 2-3 orders of magnitude smaller than the MSD values. (d) Original trajectory (connected black points) along with tracked

data (tan dashed line) and SPIFF corrected results (tan solid line), from synthesized images with mild intensity filtering. Details of synthetic data generation and the filtering done are given in the Supporting Information. (e) MSDs obtained from tracked images that were synthesized from simulated, high frame rate trajectories. For comparison, the pink stars are the first two experimentally measured MSD values shown in panel (a) Inset shows representative synthetic image (scale bar represents 400 nm). (f) Difference between original and SPIFF-corrected MSD for synthetic data with different radii (*R=1,2,3*). Note that the difference is always positive, implying that the error in the MSD is consistently reduced, and that the values of the differences are significantly greater than in panel (c).

To test this interpretation, we generated synthetic particle trajectories based on the transport properties given by Lin et al[33], used them to generate particle positions and images at higher frame rates (2000 FPS) and tracked them using Mosaic (see Supporting Information). Similar to the results for nanoparticles, we observe errors in the tracked trajectories and MSD values compared to the true (original) values. As can be seen from Figure 5(d-f) these errors are reduced by SPIFF correction, and the improvement is greatest for a window with a radius of *R=1*. The fact that the SPIFF correction did not reduce the MSD error in the original low frame rate video can be understood as the particle motion being undersampled for efficacious SPIFF correction; i.e. particle displacements between consecutive frames were far larger than the magnitude of the sub-pixel SPIFF corrections. This was confirmed when we reduced the frame rate of our synthesized data (see Supporting Information).

**Discussion and conclusions:** We have demonstrated the pixel-locking error that is inherent to tracking of nanoscale objects such as nanoparticles, quantum dots and fluorescent single molecules. We showed that this error can be corrected using the SPIFF algorithm, resulting in a marked improvement in the fidelity of the calculated results to the true values. Since the imaging and tracking of nanoscale objects is commonplace in contemporary research, the present work makes clear the ubiquity of tracking errors that can be particularly insidious for such localization studies. The construction of the meta-pixel SPIFF distribution and correction presents a general and straightforward way to identify and correct such errors.

Placed in a broader perspective, SPIFF correction of particle tracking errors is especially valuable when the Nyqiust sampling condition is compromised in both time and space in the experimental data. When the tracked particles are small (or when the data is crowded) and the sampling rate is sufficiently large that particle displacement between frames is sub-pixel, then the SPIFF correction significantly improves the measurement (localization) accuracy. Greater magnification cannot simply solve the issue given the diffraction limit and Rayleigh resolution criterion and the tradeoff between precision and accuracy that occurs when spreading a photon-limited source over more pixels. Therefore, the importance of this analysis and SPIFF correction will grow as researchers continue to improve the resolution and rate of imaging techniques[35–38], and demand greater accuracy.

**Acknowledgements**: We thank Prof. Aaron Dinner for stimulating comments. The authors would like to acknowledge support from the Vannevar Bush Faculty Fellowship program sponsored by the Basic Research Office of the Assistant Secretary of Defense for Research and Engineering and funded by the Office of Naval Research through grant N00014-16-1-2502. The ED-LD simulations were performed at the Center for Nanoscale Materials, a U.S. Department of Energy Office of Science, User Facility under Contract No. DE-AC02-06CH11357.

# Supporting Information for: Analysis and correction of errors in nanoscale particle tracking using the Single-pixel interior filling function (SPIFF) algorithm.

Yuval Yifat[1], Nishant Sule[1], Yihan Lin[3], Norbert F. Scherer[1,2,*]

[1] *James Franck Institute,* [2] *Department of Chemistry, The University of Chicago, Chicago Il. 60637, USA.*

[3] *Center for Quantitative Biology, Peking-Tsinghua Center for Life Sciences, Academy for Advanced Interdisciplinary Studies, Peking University, Beijing 100871, China.*

\* Corresponding Author: nfschere@uchicago.edu

**1. Nanoparticle trapping experimental setup:**

The 150 nm Ag particles described in the text were trapped in a linearly polarized Gaussian laser beam that was focused through a microscope objective. A diagram of the trapping setup along with a representative image (i.e. one frame in a video) from the experiment and measurements of the beam profile and point-spread function (PSF) of a single Ag nanoparticle are shown in Figure S1.

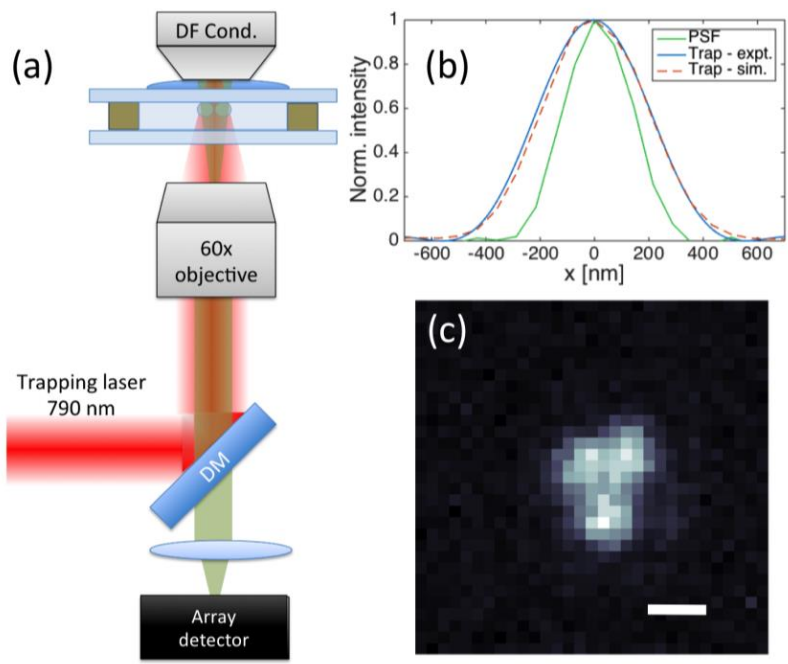

**Figure S1**. (a) Diagram of optical trapping setup used to obtain the experimental data presented in the main text. DF Cond.–Dark field condenser, DM-Dichroic mirror. (b) Plots of the measured intensity of the Gaussian optical trap and the corresponding intensity profile of the focused Gaussian beam used in the ED-LD simulations. The point-spread function (PSF) of a single Ag nanoparticle for our optical system is also shown. The optical tweezers instrument was used for the optical trapping experiments presented in the text, while the simulated focused Gaussian beam was used for the ED-LD simulations described in Fig. 3 in the main text. The FWHM for both is 450 nm, and the FWHM of the particle PSF is 300 nm. (c) Dark-field scattering image of three Ag particles trapped in the Gaussian laser trap. Scale bar is 360 nm.

The setup consisted of a CW Ti:Sapphire laser emitting linearly polarized (LP) light at a wavelength of 790 nm that was collimated and directed into a Nikon TI inverted optical microscope and through a 60x IR corrected water immersion objective (Nikon 60x Plan APO IR water immersion objective, NA = 1.27). The incident power was measured to be 8mW before the Dichroic mirror. The laser beam was focused into a sample cell that was filled with a solution of 150nm diameter silver nanoparticles coated with a layer of polyvinylpyrrolidone (PVP) diluted in 18MΩ de-ionized water in a ratio of 1:200. The beam was positioned such that its focus would be in the solution slightly below the top cover slip and had a measured FWHM of 450 nm. The particles were illuminated using a dark-field condenser and the light they scattered in a dark-field configuration was captured by the objective and imaged onto an sCMOS camera (Andor Neo; 6.5 µm pixel size) with a total magnification of 90x. The particle motion was captured with an exposure time of 0.1ms at a frame rate of 1040 frames per second.

The size of the particles on the CMOS detector, that is the number of pixels they occupy as an image, is due to the optical magnification of the system. If the magnification were too small, the particle images would occupy only a few or even a single pixel, increasing the pixel-locking error. Therefore, one might make the claim that increasing the magnification of the system (to 120x, say, as is shown for the same type of particles in ref [1]), would increase the particle size on the detector and allow the use of larger windows for particle tracking, thereby decreasing the pixel locking. However, increasing the magnification will decrease the SNR of the system, as the same number of photons will now be spread over a larger number of pixels, thereby affecting the precision of localization. This is of particular concern in the case of photon-limited experiments such as imaging of single fluorescent molecules or rapidly moving particles that necessitate short integration times. Therefore, the current magnification is typical and useful for imaging particles of this size.

## 2. Preparation and analysis of synthetic images of nanoparticles:

Since it is not possible to determine the "true" positions of the tracked particles in an optical trapping experiment of mobile particles, we devised a method for simulating the results of such an experiment and applied the tracking algorithms to this "synthetic data", thereby creating a benchmark for the performance of the SPIFF correction. The synthetic image data were created by simulating images of nanoparticles based on actual experimental images and videos (such as those described by Yan et al [2]). We analyzed the images (frames) taken from experimental data and extracted the representative background noise level and its variation, as well as the intensity and width of a typical detected Ag particle. These values were then used to parameterize a procedure that accepts an input position list $\{x_i, y_i\}_k$, which describes the coordinates of each one of *M* particles throughout *N* frames (the dimensions of the list are *2MxN*). We then generate *N* synthetic frames (images) displaying the positions of the *M* particles throughout the video. Essentially, we are creating image files, which we call synthetic images (or frames), that simulate the intensity distribution that is recorded by a CMOS (or CCD) detector in an experimental setup for a given list of particle positions. An example of experimental and synthesized frames of two 150 nm Ag particles at different separations are shown in Figure S2(a-c) and S2(d-f), respectively.

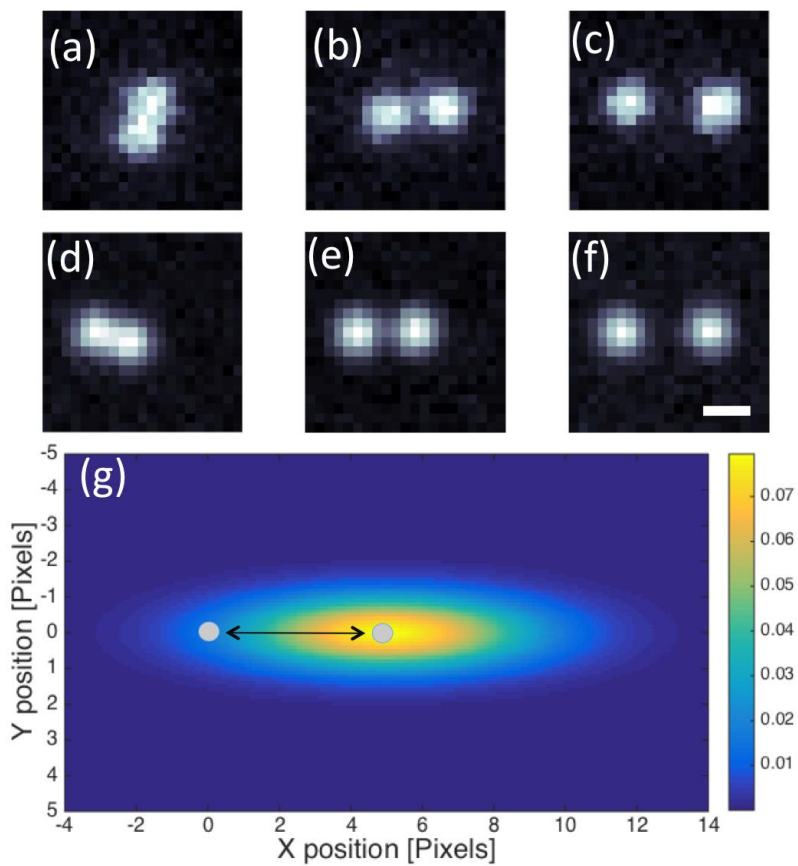

**Figure S2**. Representative experimental images of two 150nm diameter Ag particles separated by (a) 3.5, (b) 6.1, (c) 9.3 pixels respectively. (d-f) Representative synthesized images of similar particles with separations matching the experimental conditions. Scale bar is 360 nm. (g) Probability distribution for the location of particle 2 as described in the main text. Particle 1 is located at (0,0), while particle 2 is placed at $10^4$ positions around (5,0) with a normal distribution with a standard deviation of 2 and 1 pixels along the x and y axes, respectively.

Once the synthetic images were created, we used them to analyze the errors in particle localization by tracking errors in interparticle separations and angles as described in Figure 2 in the main text. This was done by simulating $10^4$ particle positions of two particles, the first was fixed at location (0,0) while the second was randomly positioned around (5,0) with a Normal distribution with a standard deviation of 2 and 1 pixels along the x and y axes, respectively. A graphical representation of the position distribution of particle 2 for this simulated data is shown in Figure S2(g). The frames, which were synthesized from this distribution, were tracked and analyzed as described in the main text. The data in Fig. S2(g) are the same as shown in Fig. 2(c).

## 3. Further improvement of the accuracy of interparticle distance determination:

Close examination of the distribution of tracked particle positions shown in Figure 2(c) in the main text and again in Fig. S3(a) reveals that when the interparticle separation becomes small, and the intensities from the two particles overlap on the detector, the identification percentage, that is the percentage of frames in which two particles are identified decreases, and the meta-pixel distributions are not evenly biased around the pixel center. The skew in the tracked localizations is demonstrated in Figure S3, which shows a single row in the distribution of localizations from the main text. As a result, the SPIFF-corrected particle positions are skewed towards the edge of the pixel in the frames in which the interparticle separation is small instead of being evenly distributed throughout the pixel. This skewing bias cannot be corrected with a single "one size fits all" SPIFF inverse function. Thus, an error remains in the SPIFF corrected data.

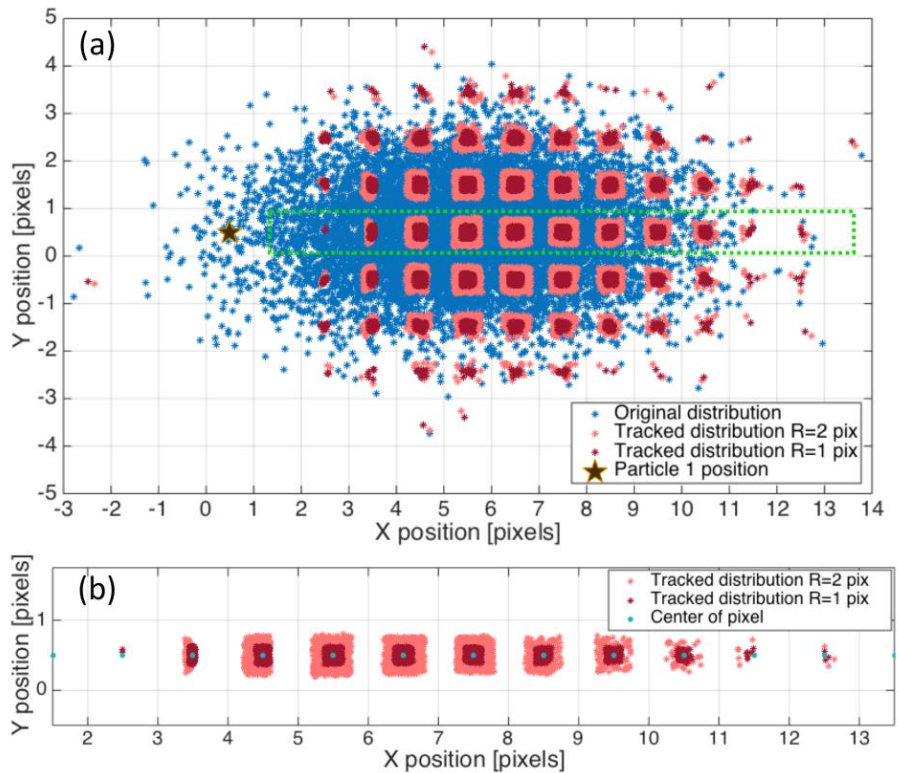

**Figure S3.** (a) Distribution of localizations for particle 2 (blue dots) in relation to the fixed position of for particle 1 (marked as black star). These distributions are the basis for analysis of pixel locking errors and SPIFF correction, as described in main text and shown in Figure 2(c). Red and pink dots show the distribution of tracked particles using the Mosaic nonlinear least-squares Gaussian fitting algorithm with a window of *R=1* and *R=2*, respectively. Only frames where both particles are identified are considered. Particle 1 was localized at (0.5,0.5) of the pixel coordinate system due to the Mosaic algorithm. The dotted green rectangle outlines a selected range highlighted in the panel (b). (b) A single row from the distribution map shown in (a) and the main text in Fig. 2(c). All localizations that were found between y = 0 and y = 1 pixels. Cyan dots designate the pixel centers. Some of the distributions localized to the pixels exhibit a skew. In this figure 1 pixel is equal to 72x72 nm in size.

Figure S3 shows that the distribution of localizations becomes skewed around the pixel center as the interparticle separation decreases. The overlap in the intensities of the two particles skews the meta-pixel distribution. This is particularly evident for particles that are localized to pixels x=2 and x=3, which are very close to the position of particle 1.

This non-uniform biasing error can easily be corrected in this case where one particle is fixed by applying the SPIFF algorithm on a local, pixel-specific meta-pixel distribution, rather than on the global distribution taken from the entire video. In doing so one needs to evaluate Eqn. (1) in the main text relative to the center of mass of the metal-pixel distribution for that specific pixel rather than relative to the center of mass of the meta-pixel distribution from the entire tracked video or relative to the pixel center. However, this solution would only work when one of the objects in the image is fixed throughout the video and therefore, the calculated position of the second particle uniquely determines the interparticle separation. However, in most experimental tracking scenarios all particles are free to move and a given interparticle separation does not correspond to a specific pixel position. Thus, the meta-pixel distribution of any given pixel will be composed of a large set of data-points originating from many different interparticle separations. As a result, the meta-pixel will not be skewed and pixel-specific correction will be ineffective.

Instead, further improvement to the localization accuracy beyond application of the SPIFF correction algorithm can be achieved by determining the separation (distance)-dependent error. This can be done by fitting the error data (i.e. the difference between the tracked particle separations and the corresponding true separation values) obtained from the analyzed synthetic frames and applying it to measurements that are performed under the same experimental conditions as the synthetic data are calculated for. As an example of this we use the simulated particle trajectories described in the main text and in Figure 3(h), calculate the mean error in separations as a function of the true interparticle separation, and fit it to a cubic function as shown in Fig S4.

Figure S4 shows the difference between the true interparticle separation, $d_{true}$, and the tracked interparticle separation, $d_{estimated}$, sorted by interparticle separation (red connected dots). We also plotted this for the SPIFF-corrected separations (black dots) and fit their distribution to a cubic function that is overlayed on the distributions (blue curve). $d_{true}$ - $d_{estimated}$ is also known as the distance error. As can be seen, the error is largest when the interparticle separation is ~5 pixels and reduces to 0 when the particles are separated by more than 7 pixels. The cubic fit shown in the figure (blue) provides an informed correction to interparticle separations taken from experimental datasets where there is no ground-truth data to compare to.

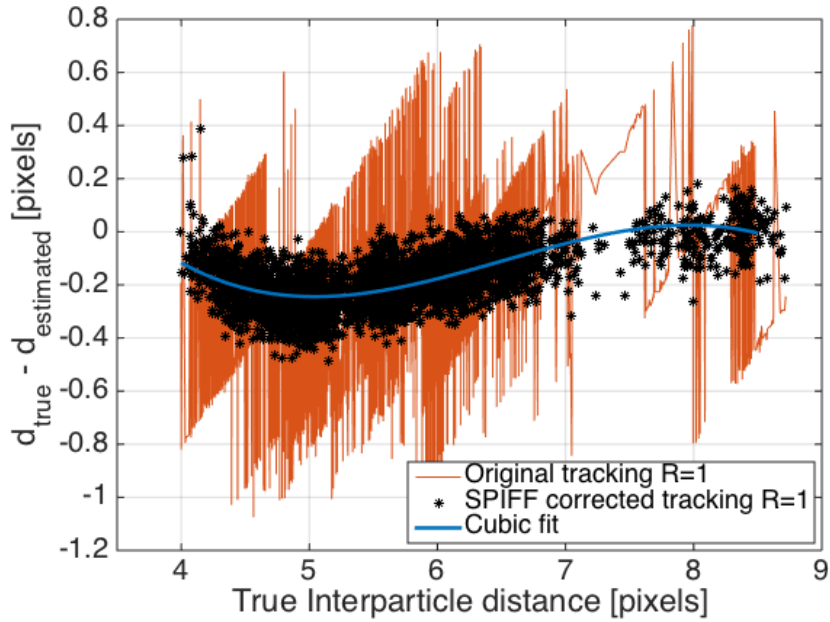

**Fig S4**. Error in interparticle separation defined as $d_{true}$ - $d_{estimated}$ for the tracked (red) and SPIFF corrected (black) synthetic data, as well as a cubic fit to the SPIFF corrected data (blue curve). See text for details.

The cubic fit (blue) provides additional information about how the imaging system skews the particle position as a function of the interparticle separation. This cubic fit can be used to further improve the accuracy of localization in experimental data. In other words, since we have based our synthetic frames, and all observations obtained from them, according to the experimental results, we can now use what we have learned from the synthetic data to improve the analysis of the experimental data. This is done by using the cubic fit shown in Fig. S4 to correct (or fit) data from both the data from ED-LD simulation and experiment. The fitted separation $d_{fitted}$ for a given frame $n$, is calculated as

$$d_{fitted}(n) = d_{estimated}(n) + g(d_{estimated}(n)), \qquad (1)$$

where $g(d)$ is the cubic fit of the distance error as a function of interparticle separation that has been extracted from the data and shown in figure S4. The results of applying the fit to the data are shown in Fig S5, which presents the time evolution of the separation as well as its distribution before and after the fitting. Where applicable, the corrected data are compared to the original data. The improvement is significant. Note that when the interparticle separation is relatively large (as seen in the left side of Fig. S5(a)) the cubic fit is small and there is a negligible difference between the SPIFF corrected and the error- fitted interparticle separation values.

The results shown in Fig. S5 are intended as a simple demonstration of the additional approach to improve the accuracy of particle tracking data. The analysis presented in this section demonstrates the potential for using the information gained from simulated data to further improve experimental results. Although the results presented here are based on our analysis of imaging of 150 nm diameter Ag particles in the optical setup described in Section 1, the approach is general and can be applied to any imaging system and tracked object. Thus, we believe that this using synthesized data, along with more sophisticated fitting methods, will improve the localization accuracy of experimental data, and warrants further research.

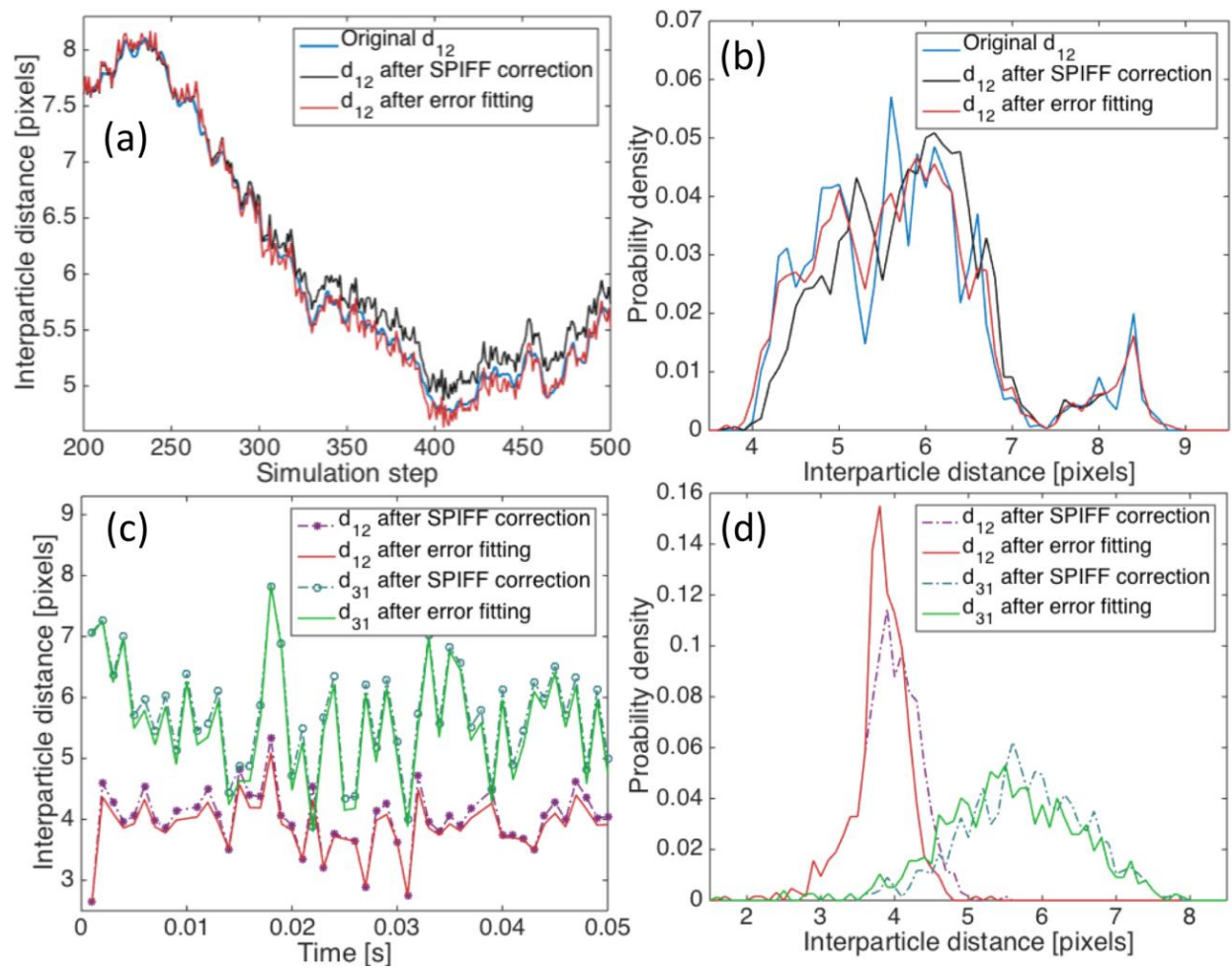

**Figure S5.** Further improvement of experimental and simulated interparticle separation measurement by application of cubic error fitted data as shown in Fig. S4. (a) Time dependent interparticle separation taken from the ED-LD simulation set described in the main text and in Figure 3. Panel (a) shows the interparticle separation calculated directly after SPIFF correction (black) and after further improvement using fitted error data (red) and how both compare to the original data (blue). Note how the fitted curve (red) improves the fidelity of the measured separation to the true data compared to the SPIFF corrected curve (black), especially for smaller interparticle separations. (b) Probability distribution of interparticle separations for the original data compared to the SPIFF corrected and fitted data. Experimental data analysis – time dependent separation (c) and probability distribution (d) for data taken from the experimental video described in Figure 4 of the main text. The definitions of the separation $d_{12}$ and $d_{31}$ are given in the main text and in Figure 4.

## 4. Preparation of trajectories and synthetic frames of single fluorescent protein tracking:

The paper by Lin et al [3] describes how single fluorescent proteins moving on a single ds-DNA strand under different flow conditions were tracked. They used the tracked data to understand and model the protein behavior by fitting their data to a drift-diffusion Langevin equation and solving it to obtain a probability density function (PDF) that describes the distribution of step sizes for a flow-biased walk:

$$P(x,t) = \frac{1}{\sqrt{4\pi Dt}} exp - \left[\frac{1}{4Dt}\left(x - \frac{\alpha ut}{\gamma}\right)^2\right]. \qquad (2)$$

In this equation, *P* is the probability density of step sizes, x, and as a function of the time step, *t, u* is the flow rate, $\alpha$ and $\gamma$ are the drag coefficients and $D = kT/\gamma$ is the diffusion coefficient of the protein on DNA.

Using this equation, we generated lists of $10^4$ displacement values for the experimental flow values reported in the paper (2,3 and 4 µl/min) given a time step of 0.5ms. We then generated simulated trajectories of the protein under different flow conditions by cumulatively summing these steps (see Figure S6). This is essentially Monte-Carlo sampling of a Wiener process. By using a time step of 0.5ms, we can effectively sample the motion of the particle at 2000 FPS, a value that is two orders of magnitude faster (larger) than what was reported in the CCD-based experiments. However, this large frame rate can be achieved with a sCMOS camera in a small region of interest by cropping to reduce the portion of the array being used.

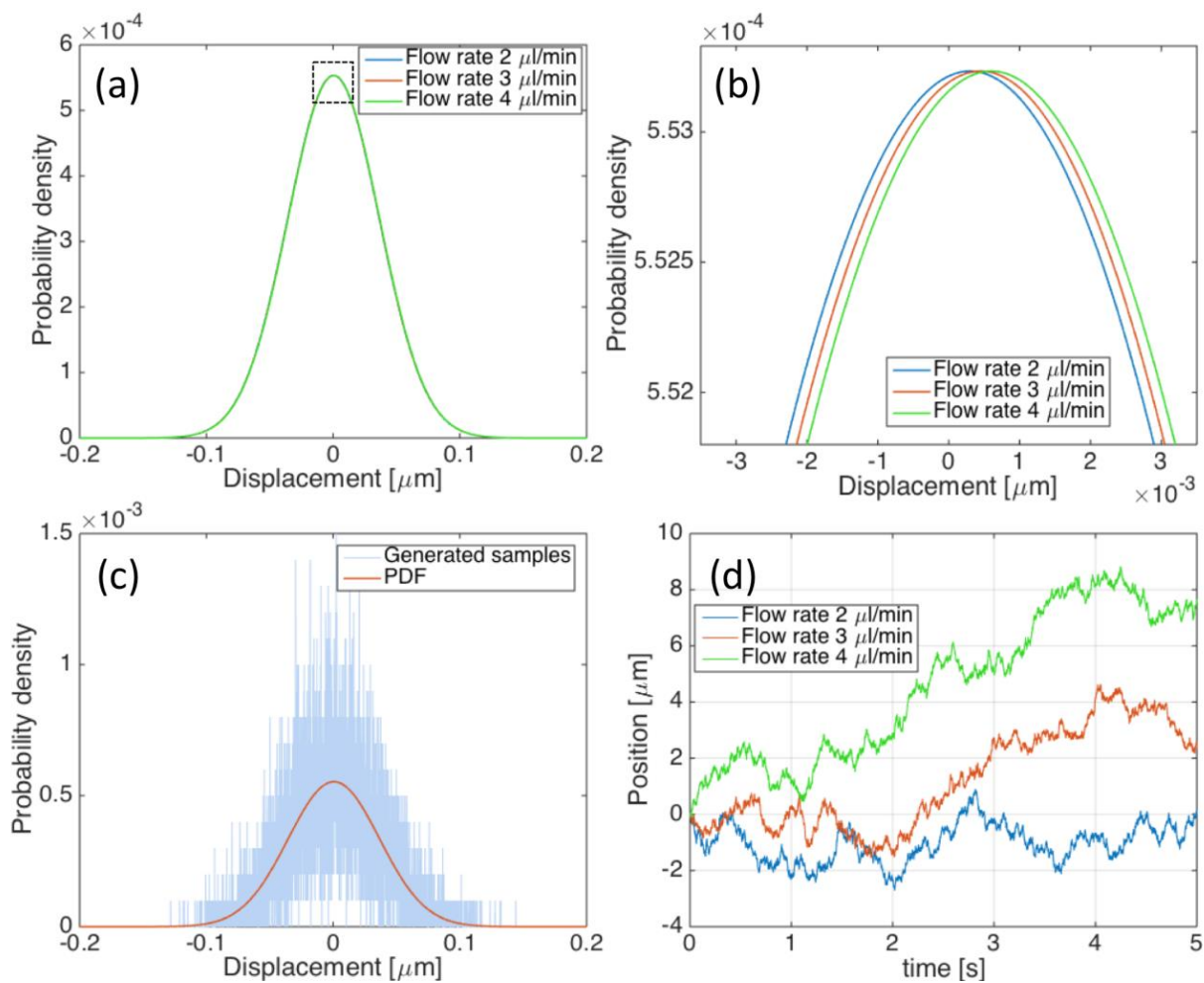

**Figure S6**. (a) Probability density function (PDF) of step sizes for a time step of 0.5ms and three different flow rates (2, 3 and 4 ul/min) used in original experiment by Lin et. Al [3]. (b) Zoomed in view of panel (a) of the region marked by the black rectangle. Note the slight difference in the positions of the maxima for different flow rates. (c) PDF for a flow of 4 ul/min (red line) along with a normalized histogram of $10^4$ random samples from the distribution (blue bars). The random samples were summed to create trajectories. (d) Particle trajectories for flow rates of 2, 3, and 4 ul/min (blue, red and green respectively). The trajectory for 4 ul/min was used to synthesize the frames for particle tracking.

These trajectories were then used to create a list of coordinates that was input into the image synthesis procedure described above. Similar to what was described in section 2 of the SI, we analyzed the experimental frames and extracted the mean intensity and standard deviation of the background pixels. These exhibited a Gaussian intensity distribution in addition to a few counts of high intensity pixels that are due to other fluorescent proteins streaking though the frame outside the plane of focus. The background of the synthetic frames was based on the Gaussian part of this distribution. We also analyzed the mean value and variance of the intensity of the particle. In addition, we fitted a 2D Gaussian function to images of the particle intensities (in pixels on the detector) to extract the mean point spread function of the experimental imaging system. We then used these extracted values to simulate the particle intensity at the desired location on the frame.

However, for most experimental frames the protein intensity distribution is not Gaussian due to blinking and fluctuations of the single molecule intensity and presumably some fast motion (see Fig S7(a)). To simulate this effect, we added an additional intensity filter that, for the pixels around the protein location, either increased or attenuated the pixel intensity by a random, normally distributed, value. This value determines the strength of the filter and acts to create a less uniform intensity map around the particle position. Examples of a frame with no additional filtering and with mild (variance of 10%) strong (variance of 25%) and extreme (variance of 50%) filtering values are shown in Figure S7. The analysis described in the main text was performed on the strongly filtered frames, except for Figure 5(e). In that panel, which shows the original and SPIFF-corrected trajectories, we chose the values of the mildly filtered frames to improve the clarity of the trajectory shown in figure 5(e).

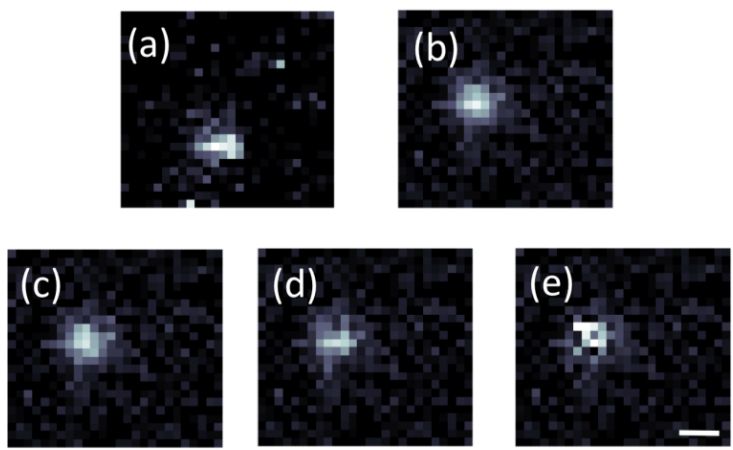

**Figure S7**. Representative images of fluorescent single molecule and synthetic images. (a) Experimental image. (b) Synthesized image with no additional filtering. (c) Synthesized image with a mild filtering (variance of 10% in the intensity of pixels associated with the particle). (d) Synthesized image with a strong filtering (variance of 25% in the intensity of pixels associated with the particle). (e) Synthesized image with extreme filtering (variance of 50% in the intensity of pixels associated with the particle). Scale bar represents 400 nm.

The application of these filters acts to replicate the noise in the experimental images. We observe pixel locking in all cases. The application of the SPIFF correction algorithm improves the fidelity of the MSD to the true values even when there is no filter applied (see Figure S8). However, Figure S8 shows that as the noise from the filtering process increases, the fidelity of the tracked trajectory to the true value decreases. Despite this, the SPIFF correction still reduces the error in the MSD by up to 0.25 pixels, demonstrating that SPIFF correction is effective even when applied to very noisy experimental image data.

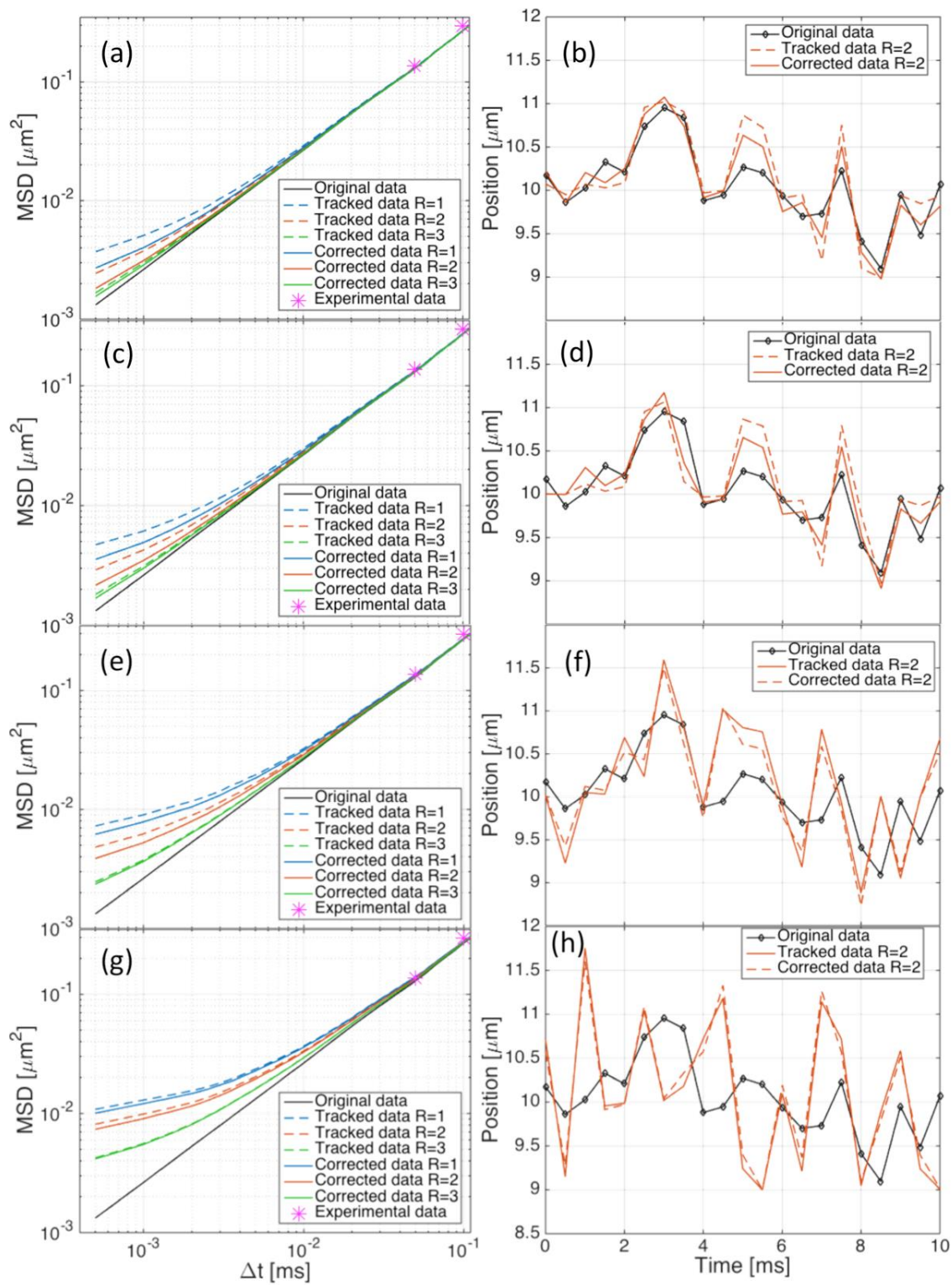

**Figure S8.** Mean Square Displacement (MSD - left column) and trajectory (right column) plots obtained from tracking frames that were synthesized under different filtering conditions. (a,b) No filtering (c,d) Mild filtering (e,f) Strong filtering (g,h) extreme filtering.

**5. The effect of decreased frame rate on fluorescent molecule trapping:**

As was mentioned in the main text, when we applied the SPIFF algorithm to the original experimental single fluorescent protein data, which was captured at 20FPS, we did not observe a consistent improvement in the MSD; i.e. the difference between the MSD calculated from the original and the corrected values varied between negative and positive values. However, when we analyzed this for the synthesized 2000FPS data, the corrected MSD values were consistently closer to the true values than those calculated from the pixel-locked tracked data. An example of the fidelity of the high frame rate data is shown in Figure 5(f), which shows that the difference between the original MSD and the corrected MSD is positive for all time separations, in contrast with Figure 5(c), which shows the experimental data and has both positive and negative values. The reason for the pronounced improvement of the SPIFF correction for high frame rate videos relative to what was seen in the experimental data is that the original data was sampled at 20FPS, and as a result, the particle displacement between frames is larger than the pixel locking error. Therefore, applying the SPIFF algorithm could either increase or decrease the MSD. If the data were sampled at a high enough frame rate, the mean particle displacement between frames will be sub-pixel and the additional sub-pixel correction obtained through the SPIFF algorithm is more likely to improve the accuracy of the MSD.

In order to verify this conjecture, we selected two subsets from the synthetic tracking data: in one subset we selected every $100^{th}$ frame and in the other we selected every $10^{th}$ frame, thus reducing the 2000FPS data to 20FPS and 200FPS respectively. The results of the MSDs calculated from these data subsets are shown in Figure S9. As expected, decreasing the sampling rate causes the mean displacement between frames to exceed 1 pixel per frame, thereby reducing the significance of the SPIFF correction. Note that for the data sampled at high frequency (Fig. S9c) the difference is always positive, which means that the SPIFF algorithm corrects the MSD value by a significant amount relative to the particle motion from frame to frame. On the other hand, for the data at lower sampling frequencies, the difference can be either positive or negative, similar to what was shown in the main text. This means that the particle motion is >1 pixel, causing the SPIFF correction to be insignificant. We referred to this latter situation as temporarily undersampled. As stated in the Conclusions, the error in accuracy can be associated with biases in tracking, but we now see that temporal undersampling results in an error that is not corrected by the SPIFF algorithm. This point requires further investigation.

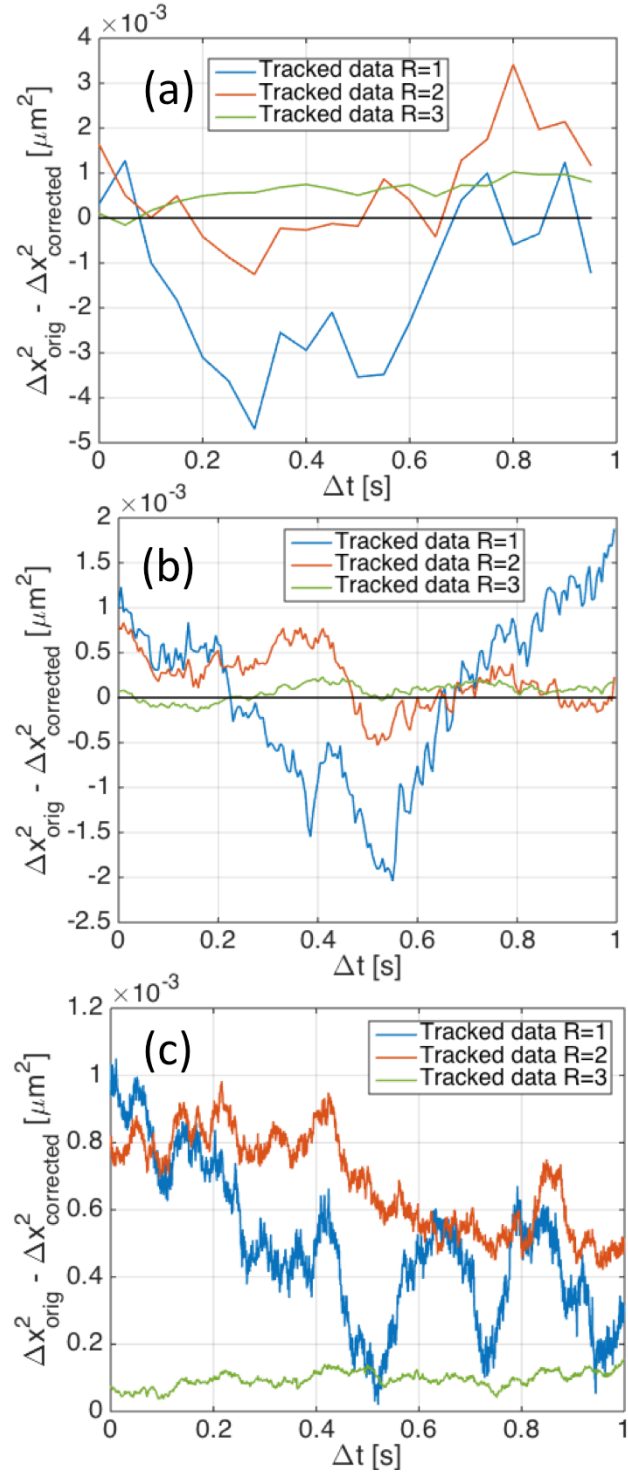

**Figure S9.** Difference between the original and the SPIFF-corrected MSD values for synthesized trajectories analyzed at (a) 20FPS (b) 200 FPS (c) 2000 FPS. Note that for the high frequency sampled data the difference is always positive, implying that the SPIFF algorithm corrects the error and brings the estimated MSD closer to the behavior expected for a Brownian random walk.

## 6. Matlab implementation of the SPIFF algorithm:

```matlab
function corrected_p = SPIFF_sort_published(p_in,SPIFF_x,SPIFF_y)

%A function that returns a SPIFF-corrected particle list based on the
%position list given in p_in, and (potentially) the x and y components
%of the SPIFF-distribution (SPIFF_xx and SPIFF_y, respectively)
%Assumption: the input p_in is a vector of size 2xn size with the
%x-coordinates in the 1st column and the y coordinates in the 2nd column.

%%%%%%%%%Generate SPIFF distribution (if necessary):%%%%%%%%%
if nargin == 0  %If the function was called empty
    corrected_p = [-1,-1];
    return
elseif nargin ==1   %No SPiff data was given
    SPIFF_x = p_in(:,1)-floor(p_in(:,1));
    SPIFF_y = p_in(:,2)-floor(p_in(:,2));
end

%%%%%%%%%%%%SPIFF correction %%%%%%%%%%%%%%%%%%%%%
%Now calculate the positive and negative probability density distributions
%(P(x_e) in the manuscript - eq. (1), and in the original manuscript
%by Burov et. al. Eq. (6)). Calculation is relative to the pixel center (0.5,0.5);
SPIFF_x_plus = SPIFF_x(SPIFF_x>=0.5);
SPIFF_y_plus = SPIFF_y(SPIFF_y>=0.5);
SPIFF_x_minus = SPIFF_x(SPIFF_x<0.5);
SPIFF_y_minus = SPIFF_y(SPIFF_y<0.5);

p_x_plus = sort(SPIFF_x_plus);
p_x_minus = sort(SPIFF_x_minus);
p_y_plus = sort(SPIFF_y_plus);
p_y_minus = sort(SPIFF_y_minus);

%The SPIFF corrected positions go here:
corrected_p = zeros(size(p_in));

%Seperates the positive and negative parts of the SPIFF in both x and y.
% for each one of the biased points, calculates the amount of points
% below its value.

for ww = 1:length(p_in)
    %Treat position list for particle 1
    if p_in(ww,1) == 0
        continue
    end
    x1_curr = p_in(ww,1);
    y1_curr = p_in(ww,2);
    SPIFF_x_curr = x1_curr - floor(x1_curr);
    SPIFF_y_curr = y1_curr - floor(y1_curr);
    if SPIFF_x_curr>=0.5
        %a. Get percentile value:
        tmp_x = sum(p_x_plus<=SPIFF_x_curr)/length(p_x_plus);
        %b. Bring back to required position:
        new_SPIFF_x = tmp_x/2;
%if the position is positive (above the floor) it stays.
        corrected_p(ww,1) = ceil(x1_curr) + new_SPIFF_x;
```

```
      else
         %a. Get percentile value (how many particles are closer to center):
         tmp_x = sum((p_x_minus)>=(SPIFF_x_curr))/length(p_x_minus);
         %b. Bring back to required position:
         new_SPIFF_x = tmp_x/2;
%If the position is negative (below the floor) it is moved by 1.
         corrected_p(ww,1) = ceil(x1_curr) - new_SPIFF_x;
      end

   if SPIFF_y_curr>=0.5
      %a. Get percentile value:
      tmp_y = sum(p_y_plus<=SPIFF_y_curr)/length(p_y_plus);
      %b. Bring back to required position:
      new_SPIFF_y = tmp_y/2;
%if the position is positive (above the floor) it stays.
      corrected_p(ww,2) = ceil(y1_curr) + new_SPIFF_y;
   else
       %a. Get percentile value (how many particles are closer to center):
       tmp_y = sum((p_y_minus)>=(SPIFF_y_curr))/length(p_y_minus);
       %b. Bring back to required position:
       new_SPIFF_y = tmp_y/2;
%If the position is negative (below the floor) it is moved by 1.
       corrected_p(ww,2) = ceil(y1_curr) - new_SPIFF_y;
   end
end

%%%%%%%%%%%%%%%%%%%%%%%%%%%%%%%%%%%%%%
```

## 7. List of videos:

**Video S1** – Synthetic movie of two Ag particles trapped in a linearly polarized Gaussian trap. Trajectories were obtained using ED-LD simulation method referenced in text, and used to synthesize movie.

**Video S2** – Experimental movie of three particles in linearly polarized Gaussian trap.

**Video S3** – Representative Experimental movie of fluorescent single molecule trajectory.

**Video S4** – Synthesized movie of simulated trajectory – no filtering.

**Video S5** – Synthesized movie of simulated trajectory – mild filtering.

**Video S6** – Synthesized movie of simulated trajectory – strong filtering.

**Video S7**– Synthesized movie of simulated trajectory – extreme filtering.